\documentstyle[aps,prb,epsf]{revtex}

\begin{document}

\draft

\title{Nonlinear Response Approach to Cooper Pair Tunneling in
Ultrasmall SIS Junctions}

\author{A. H{\"a}dicke and W. Krech}
\address{Institut f{\"u}r Festk{\"o}rperphysik,
        Friedrich-Schiller-Universit{\"a}t Jena,
        Max-Wien-Platz 1\\
        D--07743 Jena, Germany}
\date{March 1995}
\maketitle
\begin{abstract}
Single-charge tunneling in ultrasmall voltage biased SIS junctions in a
high-impedance electromagnetic environment is considered. The Cooper pair
current is calculated at $T=0$ on the basis of the elementary tunnel
Hamiltonian for quasiparticles. Therefore, the transfer of Cooper pairs
emerges automatically as an effect of higher order perturbation theory.
The supercurrent also depends on the dissipative part of the Josephson
current amplitude.
\end{abstract}
%
\section{Introduction}
%
Effects of single-charge tunneling in ultrasmall capacitance
junctions has become of much experimental and theoretical interest. For a
review see for instance Ref.\ \onlinecite{gra2}. Modern nanolithography
allows the fabrication of junctions with capacitances $C<10^{-16}$F
where the electrostatic energy differences dominate thermal fluctuations
at the 1K scale. This opens the door to a new kind of electronics.

The current through a SIS junction (see Fig.\ \ref{fig1}) can be carried by
quasiparticles (quasiparticle current $\langle I\rangle_{qp}$) and by
Cooper pairs (supercurrent $\langle I\rangle_s$).
Here we develop a nonlinear response approach to the
Cooper pair tunneling in voltage biased ultrasmall SIS junctions
basing on the elementary tunneling Hamiltonian for quasiparticles. Since
this Hamiltonian describes only the tunneling of single quasiparticles (1e)
the Cooper pair tunneling (2e) corresponds in perturbation theory to a
process of higher order. In leading order this approach yields the known
result of quasiparticle tunneling \cite{tin2}. Because of the voltage
biasing and due to Eq.~(\ref{ineq1}) the ordinary Josephson current which
is also an effect of first order perturbation theory does not play any role.

The current-voltage characteristic of small tunnel junctions is essentially
influenced by the external circuit. This electromagnetic
environment is able to absorb energy which is for superconducting electrodes
at zero temperature the only possibility to transfer the energy gain of the
tunneling process. Because Cooper pairs live in the condensate they
cannot absorb this energy. To
simplify matters we restrict ourselves to the limiting cases of low-
and high-resistance environments ($R_E\ll R_Q$ and $R_E\gg R_Q$) at $T=0$
where the quasiparticle currents are suppressed for voltages lower than the
thresholds $2{\mit\Delta}/e$ and $(2{\mit\Delta}+E_c)/e$ respectively.
$R_Q=h/e^2$ is the quantum resistance and $2{\mit\Delta}$ labels the
superconducting energy
gap. The additional part $E_c=\hbar\omega_c=e^2/(2C)$ corresponds to
the Coulomb energy. For single junctions is known that
the Coulomb blockade can only be observed if the junction is sufficiently
decoupled from the voltage bias by a high-resistance environment.
Beyond the thresholds Cooper pairs can break up into quasiparticles and
the tunnel current is carried mainly by quasiparticles.

Cooper pair tunneling is described in literature by using the model of
an effective Hamiltonian \cite{fal1,fal2,ing2,ing4,kre10} with the
perturbation term
\begin{equation}
\label{ham1}
H_T=E_J\cos {\mit\Psi}
\end{equation}
where the operator $\exp(\pm i{\mit\Psi})$ changes the macroscopic charge
$Q$ on the junction by the value $\pm 2e$ corresponding to the charge of a
Cooper pair. This means that simultaneously tunneling of two electrons
(Cooper pair) is introduced by hand. Then the calculated supercurrent reads
\cite{fal1,fal2,ing2,ing4,kre10}
\begin{equation}
\label{sc1}
\langle I\rangle_s(V)=\frac{\pi e E_J^2}{\hbar^2}\left\{
P'\left(\frac{2eV}{\hbar}\right)-
P'\left(-\frac{2eV}{\hbar}\right)\right\}
\end{equation}
with
\begin{equation}
\label{pf1}
P'(\omega)=\frac{1}{2\pi}\int\limits^{\infty}_{-\infty}
e^{4J(\tau)+i\omega\tau}\,\mbox{d}\tau\;.
\end{equation}
The function $J(\tau)$ contains the information about the structure of the
environment ($\beta=1/(k_BT)$) \cite{dev1,gra1}
\begin{equation}
J(\tau)=\frac{1}{R_Q}\int\limits_{-\infty}^{\infty}
\frac{\mbox{Re} Z_t(\omega)}{\omega}\left\{\coth\frac{\beta\hbar\omega}{2}[
\cos\omega\tau-1]-i\sin\omega\tau\right\}\mbox{d}\omega\;,
\label{jf1}
\end{equation}
where $Z_t(\omega)=1/(i\omega C+1/R_E)$. Note, that at $T=0$ one has
\[\begin{array}{lcl}
P'\left(\frac{2eV}{\hbar}\right)\Rightarrow\hbar\delta(2eV)&\mbox{for}&
\frac{R_E}{R_Q}\to 0,\\
\\
P'\left(\frac{2eV}{\hbar}\right)\Rightarrow\hbar\delta(2eV-4E_c)&\mbox{for}&
\frac{R_E}{R_Q}\to\infty.
\end{array}\]
This peak structure actually has been seen in experiment \cite{kuz1}.
The result is said to be correct if the Josephson coupling energy
$E_J=\hbar/(2e)I_c$ is much smaller than $E_c$. By use
of the known formula for the critical current $I_c$ one gets the inequality
\begin{equation}
E_c\gg \frac{R_Q}{8R}{\mit\Delta}\;.
\label{ineq1}
\end{equation}
$R$ is the normal tunnel resistance which obeys
the relation $R_Q\ll R$. Unfortunately, this model contains nearly no
information about the superconducting electrodes. In contrast to this the
quasiparticle current ($T>0$)
\begin{equation}
\label{qpc1}
\langle I\rangle_{qp}(V)=\int\limits^{\infty}_{-\infty}
\mbox{Im} I_q(\omega)P\left(\frac{eV}{\hbar}-\omega\right)
\frac{1-e^{-\beta eV}}{1-e^{-\beta\hbar\omega}}\,\mbox{d}\omega
\end{equation}
is expressed in terms of the quasiparticle current amplitude $\mbox{Im} I_q$
which depends
in a characteristic way on $\omega$.
In case of a symmetric junction the current amplitude $\mbox{Im} I_q(\omega)$
reads at $T=0$ according to standard BCS theory \cite{lik5}
\begin{equation}
\mbox{Im} I_q(\omega)=\frac{{\mit\Delta}}{eR}\left\{\begin{array}{ccc}
0&\mbox{for}&0<\frac{\hbar\omega}{{\mit\Delta}}<2\\ \\
\frac{\hbar\omega}{{\mit\Delta}}
E\left(\sqrt{1-\left(\frac{2{\mit\Delta}}{\hbar\omega}\right)^2}\right)-
\frac{2{\mit\Delta}}{\hbar\omega}
K\left(\sqrt{1-\left(\frac{2{\mit\Delta}}{\hbar\omega}\right)^2}\right)
&\mbox{for}&\frac{\hbar\omega}{{\mit\Delta}}>2
\end{array}\right.\;.
\label{caq1}
\end{equation}
The symbols $E$ and $K$ stand for the
complete elliptic integrals of the first kind \cite{grd1}.
The definition of
$P(\omega)$ differs from that of $P'(\omega)$
(Eq. (\ref{pf1})) by the lack of the factor 4 in front of the function
$J(\tau)$. The dependence of the supercurrent on the factor $I_c^2$ in Eq.
(\ref{sc1}) indicates that the supercurrent has something to do with the
squared Josephson current amplitude $\mbox{Re} I_p$. This is motivation
to express the supercurrent $\langle I\rangle_s$ by means of a
perturbation theory of higher order in the elementary tunneling
Hamiltonian
\begin{eqnarray}
H_T&=&H_++H_-\;,\qquad \qquad H_-=H_+^{\dagger}\;,\nonumber\\
\label{ham2}
H_+&=&\sum\limits_{l,r,\sigma}T_{lr}c^{\dagger}_{r,\sigma}c_{l,\sigma}
e^{i\Phi}\;,
\end{eqnarray}
where $c_{l,\sigma}$ and $c_{r,\sigma}$ stand for quasiparticle annihilation
operators of the
left and right electrode satisfying anticommutation relations.
The spin is labeled by the subscript $\sigma$.

In this way the special features of Cooper pair tunneling (transfer of
charges 2e, energy transfer only to the environment) arise automatically.
In other words we do not consider tunneling particles
with charge $2e$ from the beginning. Rather than we start with elementary
particles (electrons) with charge $1e$ and the supercurrent arises as an
effect of higher order.
Furthermore, the dependence on $\mbox{Re} I_p$ describes the transition to
a new branch if the supplied energy is able
to break Cooper pairs into quasiparticles ($eV=2{\mit\Delta}$).

Using the expression for the mean current in nonlinear
response theory those parts which correspond to the Cooper pair tunneling
can be identified and calculated in a systematic way.

\section{Response theory}
The dynamics of a physical system modeled by $H=H_o+H_T$
will be described by the statistical operator $\rho$ satisfying
the von Neumann equation which is in the interaction representation
(superscript $(I)$) equivalent to the integral equation
\begin{equation}
\label{neq1}
\rho(t)^{(I)}=\rho_o-\frac{i}{\hbar}\int\limits_{-\infty}^{t}
[H_T^{(I)}(t'),\rho^{(I)}(t')]\mbox{d} t'\;.
\end{equation}
$H_o$ is the unperturbed part of the Hamiltonian whereas the
interaction part $H_T$ reads in the interaction representation
\[
H_T^{(I)}(t)=e^{\frac{i}{\hbar}H_o\cdot t}H_Te^{-\frac{i}{\hbar}H_o\cdot t}\;.
\]
It is assumed that the interaction is switched on at $t=-\infty$
adiabatically. The operator $\rho_o$ is given by the canonical expression
\[
\rho_o=\frac{e^{-\beta H_o}}{\mbox{tr}\{e^{-\beta H_o}\}}\;.
\]
The solution can be found by successive approximation. Now the mean value
of the current operator reads
\begin{eqnarray}
\langle I\rangle&=&
\frac{1}{i\hbar}\int\limits_{-\infty}^{t}\mbox{d} t'\,
\langle[I^{(I)}(t),H_T^{(I)}(t')]\rangle_o\nonumber\\
\label{sc2}
&&+\left(\frac{1}{i\hbar}\right)^3\int\limits_{-\infty}^{t}\mbox{d} t'
\int\limits_{-\infty}^{t'}\mbox{d} t'\int\limits_{-\infty}^{t''}\mbox{d} t'''\,
\langle[[[I^{(I)}(t),H_T^{(I)}(t')],H_T^{(I)}(t'')],H_T^{(I)}(t''')]\rangle_o
+\ldots
\end{eqnarray}
In this equation the mean values have to be calculated with respect to
$\rho_o$. A term of zeroth order is missing because in case of no
interaction (tunneling) there is also no current. The first term corresponds
to linear response theory and leads either to the known quasiparticle
tunneling (Eq. (\ref{qpc1})) or to the Josephson current. The second term
describes the first nonlinear corrections.

\section{The model}
Now let us apply this theory to tunneling through a SIS junction with
environment (Fig.\ \ref{fig1}). The total Hamiltonian reads
\begin{equation}
\label{ham3}
H = H_o+H_T=QV+H_{res}+H_T\;,
\end{equation}
where the tunnel Hamiltonian is given by Eq.~(\ref{ham2}). In case of
superconducting electrodes one can assume that the macroscopic phase is
already contained in the phase operator ${\mit\Phi}$ \cite{rog1}.
Owing to this phase operator tunneling is connected with excitations
in the electromagnetic environment. The operator $H_+$ means e.g.
tunneling from left to right in contrast
to the Hermitian conjugate which describes the reverse process.
$T_{lr}$ are the tunneling matrix elements.
The basic algebra ruling this approach is the following relation
\cite{ave1}
\begin{equation}
\label{alg1}
H_{\pm}\cdot F(Q)=F(Q\pm e)\cdot H_{\pm}\:,
\end{equation}
where $F$ is an arbitrary function of the junction charge
$Q$. This algebra corresponds to the elementary commutation relation
\begin{equation}
[Q,{\mit\Phi}]=ie\;.
\label{alg2}
\end{equation}
The convention is chosen in such a way that a positive voltage
favors tunneling from left to right which reduces the junction charge
$Q$ by $e$.

The reservoir Hamiltonian $H_{res}$ consists of terms corresponding to
the left and right electrodes and the environment which can be described
in standard way \cite{dev1,gra1}.

For the calculation of the stationary mean current in terms of
Eq.~(\ref{sc2}) a current operator has to be defined. This is done
in the following way
\begin{equation}
\label{cop1}
I=-\frac{\mbox{d}}{\mbox{d} t}Q=-\frac{1}{i\hbar}[Q,H]
=\frac{e}{i\hbar}(H_+-H_-)\;.
\end{equation}
Now the quasiparticle current which is the first order term in
Eq.~(\ref{sc2}) reads
\begin{equation}
\label{qpc2}
\langle I\rangle_{qp}=-\frac{2e}{\hbar^2}\mbox{Re}\int\limits_{-\infty}^t
\mbox{d} t'\,\langle[H_+^{(I)}(t),H_-^{(I)}(t')]\rangle_o\;.
\end{equation}
The Cooper pair current is contained in the following terms of second
nonvanishing order
\begin{eqnarray}
\langle I\rangle_s=\frac{2e}{\hbar^4}\mbox{Re}\int\limits_{-\infty}^t
\mbox{d} t'\int\limits_{-\infty}^{t'}\mbox{d} t''\int\limits_{-\infty}^{t''}
\mbox{d} t'''&\Big\{&
\langle[[[H_+^{(I)}(t),H_+^{(I)}(t')],H_-^{(I)}(t'')],H_-^{(I)}(t''')]
\rangle_o\nonumber\\
&&+[[[H_+^{(I)}(t),H_-^{(I)}(t')],H_+^{(I)}(t'')],H_-^{(I)}(t''')]
\rangle_o\nonumber\\
\label{sc3}
&&+[[[H_+^{(I)}(t),H_-^{(I)}(t')],H_-^{(I)}(t'')],H_+^{(I)}(t''')]
\rangle_o\Big\}\;.
\end{eqnarray}
An explanation should be given why only correlations with vanishing
signature ($++--$ and their permutations) are taken into account.
The reason is the following. The separation of the voltage
dependence in the correlation functions ($\langle\ldots\rangle_o$) without
signature zero (e.g. $\langle H_+H_+H_+H_+\rangle_o$) leads to expressions
containing time dependent ($t$) terms. Furthermore, a phase $\phi_o$ remains
indeterminated additionally. This corresponds to common Josephson physics
where one has contributions proportional to $\sin(2eVt/\hbar+\phi_o)$ and
$\cos(2eVt/\hbar+\phi_o)$\cite{rog1}. However, in Josephson physics the phase
$\phi_o$ is defined by current biasing. Therefore, in our case of
voltage-biasing one has to average with respect to this phase factor
$\phi_o$ which yields zero.

By splitting-off the voltage dependence by means of Eq.~(\ref{alg1}) and
using new time variables
\[
\tau\equiv t-t';\qquad\tau'\equiv t'-t'';\qquad\tau''\equiv t''-t'''
\]
Eq.~(\ref{sc3}) reads
\begin{eqnarray}
\langle I\rangle_s=\frac{2e}{\hbar^4}\mbox{Re}\int\limits_0^{\infty}
\mbox{d}\tau
\int\limits_0^{\infty}\mbox{d}\tau'\int\limits_0^{\infty}\mbox{d}\tau''&
\Big\{&
e^{-\frac{i}{\hbar}eV(\tau+2\tau'+\tau'')}\kappa_1(\tau,\tau',\tau'')
\nonumber\\
&&+e^{-\frac{i}{\hbar}eV(\tau+\tau'')}\kappa_2(\tau,\tau',\tau'')
\nonumber\\
\label{sc4}
&&+e^{-\frac{i}{\hbar}eV(\tau-\tau'')}\kappa_3(\tau,\tau',\tau'')\Big\}\;.
\end{eqnarray}
\section{Calculation of correlation functions}
Now one has to deal with the three correlation functions $\kappa_1$,
$\kappa_2$ and $\kappa_3$. The operators $H_{\pm}(t)$ in the interaction
representation can be written as $\tilde{H}_{\pm}(t)\exp(\pm i{\mit\Phi}(t))$
where the operators $\tilde{H}_{\pm}(t)$ carry only the time dependence with
respect to the electrodes and the phase dependend operators
$\exp(\pm i{\mit\Phi}(t))$ carry
those with respect to the environment.
For instance the function $\kappa_1$ reads therefore
\begin{eqnarray}
\kappa_1&=&\langle\tilde{H}_+(t)\tilde{H}_+(t-\tau)\tilde{H}_-(t-\tau-\tau')
\tilde{H}_-(t-\tau-\tau'-\tau'')\rangle_o\nonumber\\
&&\times\langle e^{i{\mit\Phi}(t)}e^{i{\mit\Phi}(t-\tau)}
e^{-i{\mit\Phi}(t-\tau-\tau')}
e^{-i{\mit\Phi}(t-\tau-\tau'-\tau'')}\rangle_o+\mbox{7 further terms}\;.
\label{kap1}
\end{eqnarray}
These other terms arise due to the resolution of the interlaced commutators.
The decisive step of the identification of the contributions which
describe Cooper pair tunneling is to reduce the 4-correlators with respect to
$\tilde{H}$ into 2-correlators containing only operators with the same
signature, namely
\[
\langle\tilde{H}_+(t_1)\tilde{H}_+(t_2)\tilde{H}_-(t_3)\tilde{H}_-(t_4)
\rangle_o=\langle\tilde{H}_+(t_1)\tilde{H}_+(t_2)\rangle_o\langle
\tilde{H}_-(t_3)\tilde{H}_-(t_4)\rangle_o\;.
\]
This decomposition guarantees that only condensate states corresponding to
Cooper pairs are taken into account. One can prove this from the point of
view of the elementary operators $c^{\dagger}_{l,r}$ and $c^{\dagger}_{l,r}$.
Then the decomposition is equivalent to
\[\langle c_r^{\dagger}(t_1)c_r^{\dagger}(t_3)\rangle_o
\langle c_r(t_2)c_r(t_4)\rangle_o
\langle c_l(t_1)c_l(t_3)\rangle_o
\langle c_l^{\dagger}(t_2)c_l^{\dagger}(t_4)\rangle_o\]
and one can see that on both banks of the junction only the condensate
properties contribute. The terms which have been neglected in this
decomposition belong to quasiparticle tunneling of higher order and
processes including both quasiparticles and Cooper pairs.
The correlators $\langle\tilde{H}_+(t_1)\tilde{H}_+(t_2)\rangle_o$
can be expressed by the current amplitude $\mbox{Im} I_p$ in the following way
\cite{ave1}
\begin{eqnarray}
\kappa_{\pm}(\tau)&\equiv&\langle\tilde{H}_{\pm}(t)\tilde{H}_{\pm}(t-\tau)
\rangle_o=\langle\tilde{H}_{\pm}(\tau)\tilde{H}_{\pm}(0)\rangle_o\nonumber\\
\label{kap2}
&=&-\frac{\hbar^2}{2\pi e}\int\limits_{-\infty}^{\infty}\mbox{d}\omega\,
\mbox{Im} I_p(\omega)e^{-i\omega\tau}\frac{1}{1-e^{-\hbar\omega/(k_bT)}}\;.
\end{eqnarray}
Note the symmetry $\kappa_+(\tau)=\kappa_-(\tau)$.
The phase correlations can also be calculated, for instance by generalizing
the method presented in Ref.\ \onlinecite{ing2}.

Now the correlation function $\kappa_1(\tau,\tau',\tau'')$ can be
written as
\begin{eqnarray}
\kappa_1(\tau,\tau',\tau'')=
&&\kappa_+(\tau)\kappa_-(\tau'')e^{J(\tau+4\tau'+\tau'')}
-\kappa_+(-\tau)\kappa_-(-\tau'')e^{J(-\tau-4\tau'-\tau'')}\nonumber\\
&&-\kappa_+(-\tau)\kappa_-(\tau'')e^{J(3\tau+4\tau'+\tau'')}
+\kappa_+(\tau)\kappa_-(-\tau'')e^{J(-3\tau-4\tau'-\tau'')}\nonumber\\
&&-\kappa_+(\tau)\kappa_-(\tau'')e^{J(-\tau+\tau'')}
+\kappa_+(-\tau)\kappa_-(-\tau'')e^{J(\tau-\tau'')}\nonumber\\
&&+\kappa_+(-\tau)\kappa_-(\tau'')e^{J(\tau+\tau'')}
-\kappa_+(\tau)\kappa_-(-\tau'')e^{J(-\tau-\tau'')}\;.
\label{kap3}
\end{eqnarray}
For the environment function $J(\tau)$ see Eq.~(\ref{jf1}). We
remind of the property $J(\tau)=-i\omega_c\tau$ at $T=0$ and for
$R_E/R_Q\to\infty$.

\section{Supercurrent}
Now the contribution to $\langle I\rangle_s$ coming of $\kappa_1$
($\langle I\rangle^{(\kappa_1)}_s$) can
be calculated. Using the definition
\begin{equation}
\label{f1}
f(\omega)\equiv\frac{\mbox{Im} I_p(\omega)}{1-e^{-\beta\hbar\omega}}
\stackrel{T\to 0}{\longrightarrow}\mbox{Im} I_p(\omega)\Theta(\omega)\;,
\end{equation}
where $\Theta$ is the unit step function as well as the definition of the
function $\delta_+$
\[
\delta_+(x)\equiv\frac{1}{2\pi}\int\limits_0^{\infty}e^{ikx}\mbox{d} k\;,
\]
the result can be written as
\begin{eqnarray}
\langle I\rangle^{(\kappa_1)}_s=\frac{2\pi}{\hbar}\mbox{Re}
\int\limits_{-\infty}^{\infty}\mbox{d}\omega\int\limits_{-\infty}^{\infty}
\mbox{d}\omega'
\,f(\omega)f(\omega')&\Big\{&
\delta_+(-v-\omega-\omega_c)\delta_+(-v-\omega'-\omega_c)
\delta_+(-v-2\omega_c)\nonumber\\
&-&\delta_+(-v+\omega+\omega_c)\delta_+(-v+\omega'+\omega_c)
\delta_+(-v+2\omega_c)\nonumber\\
&-&\delta_+(-v+\omega-3\omega_c)\delta_+(-v-\omega'-\omega_c)
\delta_+(-v-2\omega_c)\nonumber\\
&+&\delta_+(-v-\omega+3\omega_c)\delta_+(-v+\omega'+\omega_c)
\delta_+(-v+2\omega_c)\nonumber\\
&-&\delta_+(-v-\omega+\omega_c)\delta_+(-v-\omega'-\omega_c)
\delta_+(-v)\nonumber\\
&+&\delta_+(-v+\omega-\omega_c)\delta_+(-v+\omega'+\omega_c)
\delta_+(-v)\nonumber\\
&+&\delta_+(-v+\omega-\omega_c)\delta_+(-v-\omega'-\omega_c)
\delta_+(-v)\nonumber\\
\label{sc5}
&-&\delta_+(-v-\omega+\omega_c)\delta_+(-v+\omega'+\omega_c)
\delta_+(-v)\Big\}\;.
\end{eqnarray}
Here, the variable $v=eV/\hbar$ is used.
The same procedure has to be employed with respect to the terms including the
other correlation functions $\kappa_2(\tau,\tau',\tau'')$ and
$\kappa_3(\tau,\tau',\tau'')$. The real part of the sum of these terms can be
calculated by means of the Dirac formula
\begin{equation}
\label{dc1}
\delta_+(x)=\frac{1}{2}\left(\delta(x)+{\cal P}\frac{1}{x}\right).
\end{equation}
Finally, at least one integration (e.g. with respect to $\omega'$, see
Eq.~(\ref{sc5})) can be carried out and one gets after a lenghty calculation

\begin{eqnarray}
\langle I\rangle_s(v)&=&\frac{\pi}{2e}\left\{\left[\frac{1}{\pi}
\int\limits_{-\infty}^{\infty}\rule{-3.4ex}{0ex}-\mbox{d}\omega\,
\frac{f(\omega)}{\omega-\omega_c}
\right]^2+f(\omega_c)\right\}\delta(v-2\omega_c)\nonumber\\
&&-\frac{f(v-\omega_c)}{2\pi e}\frac{2\omega_c}{v(v-2\omega_c)}
\int\limits_{-\infty}^{\infty}\rule{-3.4ex}{0ex}-\mbox{d}\omega\,
\frac{f(\omega)(\omega+\omega_c)}{(\omega+v-\omega_c)(\omega-v+\omega_c)}
\nonumber\\
&&-[v\to-v]\;.
\label{sc6}
\end{eqnarray}
The dash in the integral sign means that one has to take the principal value
of the integral. Eq.~(\ref{sc6}) is our main result.

\section{Discussion}
Concerning the structure of the current let us concentrate on the case of a
high-resistance environment ($\omega_c\neq 0$). The opposite case of
a low-resistance environment corresponds to the condition $\omega_c\equiv 0$.
One can make the following remarks:
\begin{itemize}
\item[(i)] The current is an antisymmetric function of the applied voltage
which reflects the expectation that a reversed voltage leads to a reversed
current.
\item[(ii)] The current shows a $\delta$-like singularity at $2eV=4E_c$
corresponding to the fact that the energy $2eV$ connected with the tunneling
of a Cooper pair has to be transferred to the environment. Because Cooper
pairs live in the condensate they cannot absorb this energy. Of course,
this singular expression will be smoothed both due to finite temperatures
and finite environment resistances.
\item[(iii)] There is an additional current contribution which is proportional
to $f(v-\omega_c)$. Because of Eq.~(\ref{f1}) and of the known structure
of $\mbox{Im} I_p(\omega)$ in standard BCS theory
($\mbox{Im} I_p(\omega)=-\mbox{Im} I_p(-\omega)$)\cite{lik5}
\begin{equation}
\mbox{Im} I_p(\omega)=\frac{2}{\pi}I_c\left\{\begin{array}{ccc}
0&\mbox{for}&0<\frac{\hbar\omega}{{\mit\Delta}}<2\\ \\
\frac{2{\mit\Delta}}{\hbar\omega}
K\left(\sqrt{1-\left(\frac{2{\mit\Delta}}{\hbar\omega}\right)^2}\right)
&\mbox{for}&\frac{\hbar\omega}{{\mit\Delta}}>2
\end{array}\right.
\label{cap1}
\end{equation}
this current contribution only exists if
$v-\omega_c\le 2{\mit\Delta}/\hbar$ or $eV\le 2{\mit\Delta}+E_c$ which is
just the condition for the onset of the quasiparticle current.
Since our approach is based on
higher order perturbation theory we are only interested in effects which
occur in the gap region of quasiparticle tunneling. Therefore, only the
first term of Eq.~(\ref{sc6}) has to be considered.
\end{itemize}
The integral in the first term of Eq.~(\ref{sc6}) reminds of the
definition of $\mbox{Re} I_p(\omega_c)$ according to the Kramers-Kronig
relation. The only difference is the $\Theta$-function in the integrand.
Nevertheless, it is reasonable to discuss the case
$\omega_c<2{\mit\Delta}/\hbar$
corresponding to the subgap region between $0$ and the position of the
Riedel peak. This means $f(\omega_c)\equiv 0$ and the
supercurrent reads for $0<v<2{\mit\Delta}/\hbar+\omega_c$
\begin{equation}
\langle I\rangle_s(v)=\frac{\pi}{2e}\left[\frac{1}{\pi}
\int\limits_{-\infty}^{\infty}\rule{-3.4ex}{0ex}-\mbox{d}\omega\,
\frac{f(\omega)}{\omega-\omega_c}\right]^2
\delta(v-2\omega_c)\;.
\label{sc7}
\end{equation}

Fig.\ \ref{fig2}, where we have plotted the expressions
\begin{equation}
\mbox{Re} I_p(\omega)=-\frac{1}{\pi}
\int\limits_{-\infty}^{\infty}\rule{-3.4ex}{0ex}-\mbox{d}\omega'\,
\frac{\mbox{Im} I_p(\omega')}{\omega'-\omega}
=\frac{2}{\pi}I_c\left\{\begin{array}{ccc}
K\left(\frac{\hbar\omega}{2{\mit\Delta}}\right)
&\mbox{for}&0<\frac{\hbar\omega}{{\mit\Delta}}<2\\ \\
\frac{2{\mit\Delta}}{\hbar\omega}K\left(\frac{2{\mit\Delta}}{\hbar\omega}
\right)&\mbox{for}&\frac{\hbar\omega}{{\mit\Delta}}>2
\end{array}\right.
\label{cap2}
\end{equation}
(see Ref.\ \onlinecite{lik5}) and
\begin{equation}
-\frac{1}{\pi}\int\limits_{-\infty}^{\infty}\rule{-3.4ex}{0ex}-
\mbox{d}\omega'\,
\frac{\mbox{Im} I_p(\omega')\Theta(\omega')}{\omega'-\omega}\;,
\label{cap3}
\end{equation}
respectively, shows for $0\le\omega<2{\mit\Delta}/\hbar$ that sufficiently
far from the position of the Riedel peak the approximation
\begin{equation}
-\frac{1}{\pi}\int\limits_{-\infty}^{\infty}\rule{-3.4ex}{0ex}-
\mbox{d}\omega\,
\frac{f(\omega)}{\omega-\omega_c}\approx\frac{1}{2}\mbox{Re} I_p(\omega_c)
\label{cap4}
\end{equation}
holds which becomes exact for $\omega_c\to 0$. Hence, for
$\omega_c<2{\mit\Delta}/\hbar$
one can write
\begin{equation}
-\frac{1}{\pi}\int\limits_{-\infty}^{\infty}\rule{-3.4ex}{0ex}-
\mbox{d}\omega\,
\frac{f(\omega)}{\omega-\omega_c}\approx\frac{1}{2}I_c\;.
\label{cap5}
\end{equation}

Using this approximation, the supercurrent reads
\begin{equation}
\langle I\rangle_s(v)=\frac{\pi}{8e}I_c^2\{\delta(v-2\omega_c)-
\delta(-v-2\omega_c)\}
\label{sc8}
\end{equation}
which corresponds for $T\to 0$ and $R_E/R_Q\to \infty$ exactly to the result
of Eq.~(\ref{sc1}). It has been shown that this formula is valid for
$0<eV<2{\mit\Delta}+E_c$ and $E_c\ll 2{\mit\Delta}$. There is no contradiction
to the inequality (\ref{ineq1}) because
$2{\mit\Delta}\gg E_c\gg R_Q/(8R){\mit\Delta}$
is satisfied provided that the relation $R_Q\ll R$ holds. However, this
condition is just necessary for single-charge tunneling because it
guarantees that quantum fluctuations can be neglected. Roughly speaking
(cf. (\ref{cap4})), formula (\ref{sc7}) shows indeed that the Josephson
current amplitude $\mbox{Re} I_p(\omega_c)$ determines the strength of the
$\delta$-like current peak at the $eV=2E_c$. Note, that this strength
becomes singular if the $\delta$-singularity tends to the onset position
of quasiparticle tunneling because from
$v=2\omega_c\to 2{\mit\Delta}/\hbar+\omega_c$
follows $\omega_c\to 2{\mit\Delta}/\hbar$. Eq.~(\ref{sc6}) shows that for
$T>0$ there are also current contributions depending on the dissipative
part of the Josephson current $\mbox{Im} I_p$ which describes pair transfer
processes via thermally excited quasiparticles.

In case of a finite environment resistance the substitution
$J(t)=-i\omega_ct$ in Eq.~(\ref{kap3}) is not possible. The investigation
has shown that the origin of the resulting $\delta$-function in
Eq.~(\ref{sc8}) is the integration over $\tau'$ in Eq.~(\ref{sc4}).
Therefore, this integration would indeed generate the function $P'(2eV)$
known from Eq.~(\ref{pf1}). But there are also functions $J(\tau)$ and
$J(\tau'')$ which are modifying the other integrations over $\tau$ and
$\tau''$. To sum up it can be said that in this stricter approach
the dependence on the environment is much more complicated than in the
model (\ref{ham1}) which leads to Eq.~(\ref{sc1}).

\section{Reconstruction of an effective Hamiltonian}
Using some simple assumptions we are going to reconstruct an effective
Hamiltonian which leads in first order perturbation theory to the same result
(\ref{sc8}). It can also be written as
\begin{equation}
\langle I\rangle_s(v)=\frac{I_c^2}{8e}\mbox{Re}\left\{
\int\limits_0^{\infty}\mbox{d}\tau\,e^{i(v-2\omega_c)\tau}-
\int\limits_0^{\infty}\mbox{d}\tau\,e^{i(-v-2\omega_c)\tau}\right\}\;.
\label{sc9}
\end{equation}
It will turn out that this effective Hamiltonian corresponds just to
the Hamiltonian (\ref{ham1}).

Our starting point is the Hamiltonian (\ref{ham3}) and the assumption that
the perturbation term $H_T$ can be written as
\[
H_T=H_++H_-=H e^{i{\mit\Psi}}+H e^{-i{\mit\Psi}}\;.
\]
This ansatz with real constants $H$ seems to be very likely because the factor
in front of the $\delta$-function in the supercurrent is also a constant.
${\mit\Psi}$ is a phase operator which is assumed to obey the commutation
relation
\begin{equation}
[Q,{\mit\Psi}]=ike\;,
\label{alg3}
\end{equation}
where the constant $k$ is for the time being arbitrary. One gets the algebra
\begin{equation}
H_{\pm}F(Q)=F(Q\pm ke)H_{\pm}
\label{alg4}
\end{equation}
and finds using Eq.~(\ref{cop1}) by linear response theory the mean
current
\begin{equation}
\label{sc10}
\langle I\rangle_s=-\frac{2ke}{\hbar^2}\mbox{Re}\int\limits_{-\infty}^t
\mbox{d} t'\,\langle[H_+^{(I)}(t),H_-^{(I)}(t')]\rangle_o
\end{equation}
with
\[H_{\pm}^{(I)}(t)=He^{\mp\frac{i}{\hbar}keVt}e^{\pm i{\mit\Psi}(t)}\;.\]
The time dependence of the phase operator governed by the environment
Hamiltonian can be calculated in a standard way.
Finally, Eq.~(\ref{sc10}) reads
\begin{equation}
\label{sc11}
\langle I\rangle_s=-\frac{2keH^2}{\hbar^2}\mbox{Re}\int\limits_0^{\infty}
\mbox{d}\tau\,
e^{-\frac{i}{\hbar}keV\tau}\left[e^{k^2J(\tau)}-e^{k^2J(-\tau)}\right]\;,
\end{equation}
where we already know the function $J$ from Eq.~(\ref{jf1}). By comparing
Eq.~(\ref{sc11}) with Eq.~(\ref{sc9}) in the limit $T=0$ the unknown
constants $H$ and $k$ can be determined
\begin{equation}
k=2,\qquad\qquad H=\frac{\hbar}{4e}I_c=\frac{E_J}{2}
\label{con1}
\end{equation}
which reproduce expression (\ref{ham1}) with ${\mit\Psi}=2{\mit\Phi}$. The
value $k=2$ shows that the effective Hamiltonian describes tunneling of
electron pairs (Cooper pairs).
In this way the transition to the effective model corresponds to the
transition from $[Q,{\mit\Phi}]=ie$ to $[Q,{\mit\Psi}]=i2e$. However, the
effective model does not contain dissipation because it depends instead of
$I_p(\omega)$ on $I_c$ only.

\acknowledgments
We would like to thank H.-O. M\"uller for useful discussions.
This work was supported by the Deutsche Forschungsgemeinschaft.


\bibliographystyle{prsty}

\begin{thebibliography}{10}

\bibitem{gra2}
H. Grabert and M.~H. Devoret, {\sl Single Charge Tunneling:
Coulomb Blockade Phenomena in Nanostructures},
NATO ASI Series B: Physics, Vol. 294,  (Plenum Press, New York and
  London, 1992).

\bibitem{tin2}
M. Tinkham, {\sl Introduction to Superconductivity},
(McGraw-Hill, New York, 1976).

\bibitem{fal1}
G. Falci, V. Bubanja and G. Sch{\"o}n, Z. Phys. B {\bf 85}, 451 (1991).

\bibitem{fal2}
G. Falci, V. Bubanja and G. Sch{\"o}n, Europhys. Lett. {\bf 16}, 109 (1991).

\bibitem{ing2}
G.-L. Ingold and Yu.~V. Nazarov, Charge tunneling rates in ultrasmall
junctions, in Ref.\ \onlinecite{gra2}, pp.\ 21--107.

\bibitem{ing4}
G.-L. Ingold, H. Grabert and U. Eberhardt, Cooper pair current through
ultrasmall josephson junctions, (preprint 1994).

\bibitem{kre10}
W. Krech and A. H{\"a}dicke, phys. stat. sol. (b)  {\bf 180}, 175 (1993).

\bibitem{dev1}
M.~H. Devoret, D. Esteve, H. Grabert, G.-L. Ingold, H. Pothier and C. Urbina,
Phys. Rev. Lett. {\bf 64}, 1824 (1990).

\bibitem{gra1}
H. Grabert, G.-L. Ingold, M.~H. Devoret, D. Esteve, H. Pothier and C. Urbina,
Z. Phys. B  {\bf 84}, 143 (1991).

\bibitem{kuz1}
L.~S. Kuzmin, Yu.~V. Nazarov, D.~B. Haviland, P. Delsing and T. Claeson,
Phys. Rev. Lett. {\bf 67}, 1161 (1991).

\bibitem{lik5}
K.~K. Likharev, {\sl Dynamics of Josephson Junctions and Circuits},
(Gordon and Breach, New York, 1986).

\bibitem{rog1}
D. Rogovin and D.~J. Scalapino,
Ann. Phys. (N.Y.)  {\bf 86}, 1 (1974).

\bibitem{ave1}
D.~V. Averin and K.~K. Likharev,
J. Low Temp. Phys.  {\bf 62}, 345 (1986).

\bibitem{grd1}
I.~S. Gradstein and I.~M. Ryshik, {\sl Tables of Series, Products, and
Integrals}, (Harri Deutsch, Thun, 1981).

\end{thebibliography}

\begin{figure}
\caption{Scheme of the circuit}
\label{fig1}
\end{figure}

\begin{figure}
\caption{Plot of the expressions according to Eq.~(\protect\ref{cap2}) (top)
and (\protect\ref{cap3}) (bottom) in units of $I_c$ versus $\omega$ in units
of ${\mit\Delta}/\hbar$ in the subgap region ($0<\omega<2{\mit\Delta}/\hbar$)}
\label{fig2}
\end{figure}

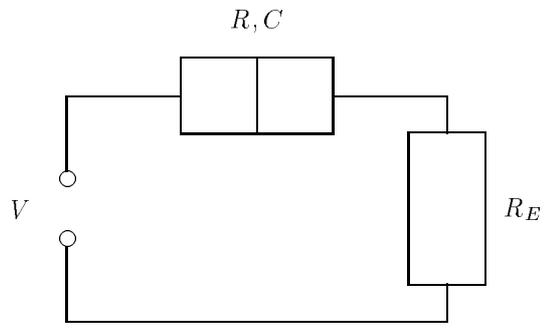
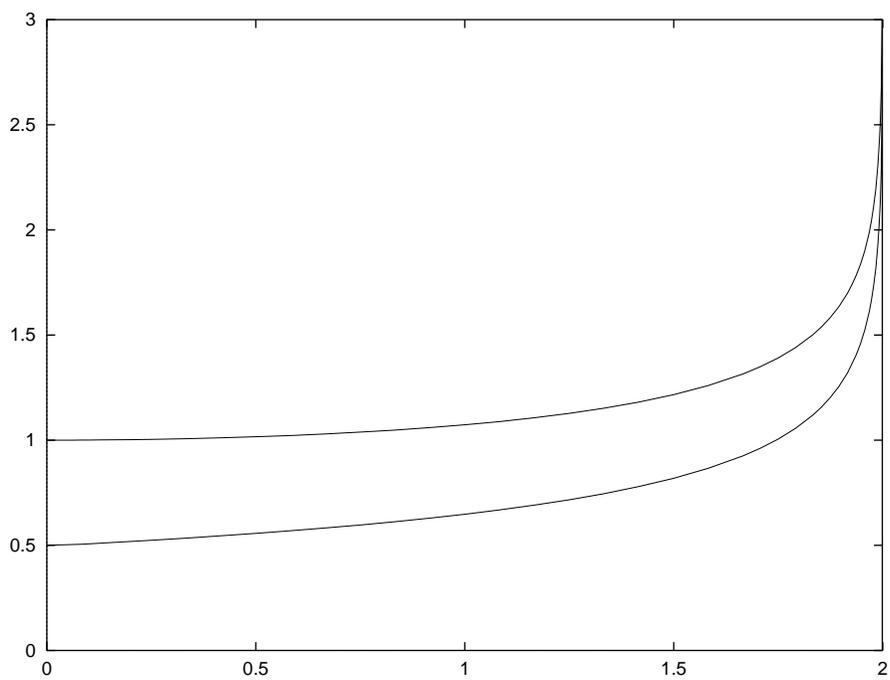
\end{document}